\documentclass{aa}

\usepackage{euclid}
\usepackage{graphicx}
\usepackage{natbib}
\usepackage{scalerel}
\usepackage{multirow}   
\usepackage{mathrsfs}
\usepackage[table]{xcolor}
\usepackage{fontawesome}
\usepackage{lipsum} 

\newcommand{\og}{{\tt OpenGADGET3}}
\newcommand{\pinocchio}{{\tt PINOCCHIO}}
\newcommand{\subfind}{{\tt SUBFIND}}

\newcommand{\pylians}{{\tt PYLIANS}}

\newcommand{\msun}{M_{\odot}\,h^{-1}}
\newcommand{\mpc}{h^{-1}\,\textup{Mpc}}

\usepackage{txfonts}
\usepackage[pdfencoding=auto,psdextra]{hyperref}
\hypersetup{
    colorlinks=true,
    linkcolor=blue,
    filecolor=magenta,      
    urlcolor=blue,
    citecolor=blue
}
\makeatletter
\renewcommand*\aa@pageof{, page \thepage{} of \pageref*{LastPage}}
\makeatother

\usepackage[utf8]{inputenc}
\usepackage{orcidlink}
\begin{document}

   \title{On the robustness of cluster clustering covariance calibration}


   \author{Alessandra~Fumagalli\orcidlink{0009-0004-0300-2535}
        \inst{1}
        \thanks{\email{a.fumagalli@physik.uni-muenchen.de}},
    Tiago~Castro\orcidlink{0000-0002-6292-3228}
        \inst{2,3,4,5}
        \thanks{\email{tiago.batalha@inaf.it}},
    Stefano~Borgani\orcidlink{0000-0001-6151-6439}, 
    \inst{6,2,3,4,5}\and
    Milena~Valentini\orcidlink{0000-0002-0796-8132}
    \inst{6,2,3,4,5}
          }

   \institute{
    $^{1}$ Ludwig-Maximilians-University, Schellingstrasse 4, 80799 Munich, Germany\\
    $^{2}$ INAF - Osservatorio Astronomico di Trieste, Via G. B. Tiepolo 11, 34143 Trieste, Italy\\
    $^{3}$ IFPU, Institute for Fundamental Physics of the Universe, via Beirut 2, 34151 Trieste, Italy\\
    $^{4}$ INFN, Sezione di Trieste, Via Valerio 2, 34127 Trieste TS, Italy\\
    $^{5}$ ICSC - Centro Nazionale di Ricerca in High Performance Computing, Big Data e Quantum Computing, Via Magnanelli 2, Bologna, Italy\\
    $^{6}$ Dipartimento di Fisica -- Sezione di Astronomia, Università di Trieste, Via Tiepolo 11, Trieste, 34131, Italy\\
             }


  \abstract{
Ongoing and upcoming wide-field surveys at different wavelengths will measure the distribution of galaxy clusters with unprecedented precision, demanding accurate models for the two-point correlation function (2PCF) covariance. In this work, we assess a semi-analytical framework for the cluster 2PCF covariance that employs three nuisance parameters to account for non-Poissonian shot noise, residual uncertainties in the halo bias model, and subleading noise terms. We calibrate these parameters on a suite of fast approximate simulations generated by \pinocchio\ as well as full $N$-body simulations from \og. We demonstrate that \pinocchio\ can reproduce the 2PCF covariance measured in \og\ at the few percent level, provided the mass functions are carefully rescaled. Resolution tests confirm that high particle counts are necessary to capture shot-noise corrections, especially at high redshifts. We perform the parameter calibration across multiple cosmological models, showing that one of the nuisance parameters, the non-Poissonian shot-noise correction $\alpha$, depends mildly on the amplitude of matter fluctuations $\sigma_8$. In contrast, the remaining two parameters, $\beta$ controlling the bias correction and $\gamma$ controlling the secondary shot-noise correction, exhibit more significant variation with redshift and halo mass. Overall, our results underscore the importance of calibrating covariance models on realistic mock catalogs that replicate the selection function of forthcoming surveys and highlight that approximate methods, when properly tuned, can effectively complement full $N$-body simulations for precision cluster cosmology.
}

    \authorrunning{Fumagalli, Castro et al.}
   \maketitle

\section{\label{sec:intro}Introduction}

Galaxy clusters, being the most massive virialized objects in the Universe, have long been recognized as powerful probes of cosmology, thanks to their sensitivity to both the geometry and growth of large-scale structure \citep[see, e.g.,][]{Allen:2011zs,Kravtsov:2012zs}. Traditional cluster studies have focused on the abundance of these systems and their variation with redshift to constrain cosmological parameters such as the matter density parameter $\Omega_{\mathrm{m}}$ and the amplitude of matter fluctuations $\sigma_{8}$~\citep{Borgani:2001,Vikhlinin:2008ym,Planck:2013lkt,Planck:2015lwi,SPT:2018njh,DES:2020cbm}. Nonetheless, the two-point correlation function (2PCF) of clusters is also a promising probe, as it is highly sensitive to the bias--mass relation and can help to break degeneracies that affect cluster abundance studies \citep{Schuecker:2002ti, Majumdar:2003mw, Sartoris:2015aga,Marulli:2018owk,Fumagalli:2023yym}. When the 2PCF is combined with other observables such as number counts or weak lensing, it can provide tighter cosmological constraints while helping to calibrate mass--observable relations \citep{Mana:2013qba,DES:2020uce,Castro:2020yes}.

An accurate assessment of cosmological constraints from cluster clustering hinges on a robust model for the covariance of the 2PCF. In contrast to the comparatively simpler case of cluster number counts \citep[e.g.,][]{Hu:2002we,Euclid:2021api}, the covariance of the 2PCF of clusters requires additional parameters to reproduce the results of numerical simulations, due to non-Gaussianities and non-linearities \citep{Euclid:2022txd}. Recent works have demonstrated that, for photometrically selected cluster surveys similar to the one anticipated from \textit{Euclid} \citep{EUCLID:2011zbd}, a Gaussian model with standard Poisson noise can fail to capture all relevant contributions in the covariance. Consequently, one must introduce ad hoc corrections, calibrated against simulations, that take into account non-Gaussian effects and the specific statistical weight of cluster samples.

In previous studies, fast approximate simulations such as those produced by the \pinocchio\ algorithm \citep{Monaco:2001jg,Monaco:2013qta,Munari:2016aut} have proven valuable for modeling the 2PCF covariance. By generating a large number of past-light-cone realizations, it is possible to calibrate the extra parameters associated with the analytical covariance to match measurements from mock cluster catalogs \citep[e.g.,][]{Fumagalli:2022plg}. In~\citet{Euclid:2022txd}, such calibrations were carried out at fixed cosmology, thus leaving the question of whether the best-fit parameters for the covariance depend on the underlying cosmological model open. Moreover, although approximate methods like \pinocchio\ can reproduce many large-scale structure statistics with remarkable efficiency, they inevitably involve simplifying assumptions in treating non-linear gravitational collapse. It is, therefore, crucial to establish whether the covariance parameters calibrated on \pinocchio\ differ significantly from those obtained from full $N$-body simulations.

Among the forthcoming wide-field surveys, the Vera C.\ Rubin Observatory's Legacy Survey of 
Space and Time (LSST)\footnote{\url{https://www.lsst.org}} \citep{LSSTScience:2009jmu}, the 
third generation of the South Pole Telescope (SPT-3G)\footnote{\url{https://astro.fnal.gov/science/cmbr/spt-3g/}} 
\citep{SPT-3G:2014dbx}, eROSITA\footnote{\url{https://www.mpe.mpg.de/eROSITA}} 
\citep{eROSITA:2020emt}, the Square Kilometre Array 
(SKA)\footnote{\url{https://www.skatelescope.org}} \citep{Maartens:2015mra}, the Dark Energy 
Spectroscopic Instrument (DESI)\footnote{\url{https://www.desi.lbl.gov}} \citep{DESI:2016fyo}, 
the Nancy Grace Roman Space Telescope\footnote{\url{https://roman.gsfc.nasa.gov}} 
\citep{Spergel:2015sza}, and \textit{Euclid}\footnote{\url{https://www.euclid-ec.org}} 
\citep{euclidoverview}, are poised to deliver measurements of summary statistics of the distribution of galaxy clusters with unprecedented 
statistical precision. These data sets will demand more accurate covariance models 
to exploit the cosmological information encoded in cluster clustering fully. 
The goal of this paper is, therefore, twofold. First, we investigate the cosmology dependence 
of the calibrated parameters that enter the covariance model for the 
cluster 2PCF. Second, we compare the parameters extracted from \pinocchio-based mocks with 
those calibrated on higher-fidelity $N$-body simulations. This analysis directly tests the 
robustness of the covariance modeling for future surveys such as \textit{Euclid}, which is 
expected to deliver a photometric cluster catalog of unprecedented size and redshift 
coverage. Ultimately, a more reliable covariance prescription will improve cosmological 
constraints from cluster clustering and deepen our understanding of systematic uncertainties.

This paper is structured as follows. In Sect.~\ref{sec:theory}, we present the theoretical framework for the 2PCF, the semi-analytical covariance model, and the procedure for fitting parameters. In Sect.~\ref{sec:methodology}, we summarize the simulations used for our covariance calibration and introduce the cluster catalogs derived from \pinocchio\ and $N$-body simulations. We compare and discuss our results in Sect.~\ref{sec:results}, where we quantify the impact of cosmology dependence and the differences between approximate and full $N$-body calibrations. We discuss the implications of our results in Sect.~\ref{sec:discussion}. Finally, we draw our conclusions in Sect.~\ref{sec:conclusions}, emphasizing the implications of our findings for the cosmological exploitation of cluster clustering in forthcoming wide-field surveys.

\section{\label{sec:theory}Covariance formalism}

This section concisely summarizes the semi-analytical model for the real-space 2PCF covariance of galaxy clusters (or halos). A more extensive derivation and validation can be found in \citet{Meiksin:1998mu} and \citet{Euclid:2022txd}, with additional details on numerical calibrations provided by \citet{Fumagalli:2022plg}.

\subsection{Baseline model and notation} 
The core of the 2PCF covariance model starts from the halo power spectrum covariance in Fourier space, which contains both Gaussian and non-Gaussian contributions \citep[see, e.g.,][]{Meiksin:1998mu,Scoccimarro:1999kp}. In real space, the halo 2PCF at a given redshift \(z\) and separation \(r\) can be expressed as
\begin{equation}
    \xi_{\mathrm{h}}(r,z) \;=\; \int \!\frac{k^2\, \mathrm{d}k}{2\pi^2}\, P_{\mathrm{h}}(k,z)\, j_0(kr)\,,
    \label{eq:halo_2pcf}
\end{equation}
where \(P_{\mathrm{h}}(k,z)\) is the halo power spectrum, and \(j_0\) is the spherical Bessel function of order zero.  On large (linear) scales, the halo power spectrum can be approximated by
\begin{equation}
    P_{\mathrm{h}}(k,z) \;=\; \overline{b}^{\,2}(z) \, P_{\mathrm{m}}(k,z) \,
    \label{eq:halo_ps}
\end{equation}
where \(\overline{b}(z)\) is the effective linear bias of halos obtained by averaging the linear bias model presented in~\citet{Euclid:2024wog} weighting it by mass according to the halo mass function~\citep[HMF;][]{Euclid:2022dbc}, and \(P_{\mathrm{m}}(k,z)\) is the linear matter power spectrum.  
When calculating the covariance, the discrete nature of halos introduces an additional source of randomness that is independent of the clustering of the matter field. This effect, known as shot-noise, is accounted for by adding a $1/\,\overline{n}(z)$ term to the power spectrum of Eq.~\eqref{eq:halo_ps}. Such a term provides the leading-order Poisson shot-noise component for a point distribution whose redshift-dependent comoving mean number density is $\overline{n}(z)$.
When transforming to configuration space and integrating over redshift bins, one obtains a baseline (Gaussian + lowest-order shot-noise) model for the 2PCF covariance \citep[cf.][]{Cohn:2005ex,Hu:2002we}:
\begin{align}
    \label{eq:cov_baseline}
    C_{ij}(z) \;=\;& \frac{2}{V_z}\, \int \frac{k^2\, \mathrm{d}k}{2\pi^2}\,\Biggl[P_{\mathrm{h}}(k,z) + \frac{1}{\overline n(z)}\Biggr]^2\,W_{i}(k)\,W_{j}(k)\,\nonumber\\
    &+ \frac{2}{V_z} \int \frac{\mathrm{d} k\,k^2}{2 \pi^2} {\overline b\,}^2 P_{\rm m}(k)\,\Biggl[\frac{1}{{\overline n}(z)}\Biggr]^2 \, W_j(k) \ \frac{\delta_{ij}}{V_i}\,.
\end{align}
Here \(V_{z}\) is the comoving volume in the redshift slice, \(W_{i}(k)\) is the window function for the \(i\)-th separation bin, and the sub-leading terms account for additional shot-noise corrections. In principle, one can include contributions from higher-order correlation terms~\citep{Meiksin:1998mu,Scoccimarro:1999kp} and from super-sample covariance~\citep{Takada:2013wfa} for more accuracy, but they often add significant complexity and require the calibration of nuisance parameters.

\subsection{Limitations of the baseline model}
Equation~\eqref{eq:cov_baseline} was shown to underestimate the total covariance when compared to simulations~\citep[see, for instance,][]{Euclid:2022txd}. Known limitations include:
\begin{itemize}
    \item Non-Poissonian Shot Noise: The simple \(1/\overline{n}\) term in general does not capture the non-Poissonian nature of the sampling of the underlying density field provided by the biased halo distribution, especially at high masses or high redshifts;
    \item Inaccurate Bias Prescription: Linear bias formulae sometimes introduce residual offsets in amplitude or scale dependence;
    \item Neglected Higher-Order Terms: Bispectrum and trispectrum contributions can be non-negligible for cluster-scale halos, particularly if the survey volume or mass threshold is such that the number density is small.
\end{itemize}
These effects can lead to a mismatch of up to tens of per cent between the baseline model and numerical simulations \citep{Euclid:2022txd}.

The halo bias and shot-noise terms in the covariance can be rescaled by empirical parameters that effectively absorb the missing (or inaccurately modeled) contributions to correct for these discrepancies. In practice, one modifies Eq.~\eqref{eq:cov_baseline} as:
\begin{align}
  C_{ij}^\mathrm{fit}(z) \;=\;& \frac{2}{V_z} \int \!\frac{k^2\,\mathrm{d}k}{2\pi^2}
  \Biggl[\bigl(\beta\, \overline{b}\bigr)^2 P_{\mathrm{m}}(k,z) + \bigl(1 + \alpha\bigr)\frac{1}{\overline{n}(z)}\Biggr]^2 
  W_i(k)\,W_j(k)\,\nonumber\\
  &+ \frac{2}{V_z} \int \frac{\mathrm{d} k\,k^2}{2 \pi^2} \bigl(\beta\, \overline{b}\bigr)^2 P_{\rm m}(k)\,\Biggl[\bigl(1 + \gamma\bigr)\frac{1}{\overline{n}(z)}\Biggr]^2 \, W_j(k) \ \frac{\delta_{ij}}{V_i}\,,
  \label{eq:cov_fit}
\end{align}
where \(\alpha\), \(\beta\), and \(\gamma\) are treated as nuisance parameters to be calibrated against a suite of mock halo catalogs extracted from simulations. In summary,
\begin{itemize}
    \item \(\beta\) corrects for residual bias errors (i.e., an over- or underestimation of \(\overline{b}\)),  
    \item \(\alpha\) adjusts the main shot-noise amplitude to account for non-Poissonian behavior and missing non-Gaussian terms, 
    \item \(\gamma\) modifies a secondary shot-noise contribution in the diagonal terms, mostly contributing at small scales.
\end{itemize}
By fitting \(\{\alpha, \beta, \gamma\}\) to simulation measurements, one can achieve better than 10 percent agreement between the analytical covariance and the fully numerical result, at least for a fixed cosmology \citep{Euclid:2022txd}. 

\subsection{Fitting procedure}
The fitting method, presented in \citet{Fumagalli:2022plg}, is designed to efficiently construct a reliable covariance matrix by constraining a model with free parameters. The procedure begins by defining a model covariance matrix, which may be incomplete or approximate, and introducing free parameters to capture unknown or uncertain contributions to the covariance. The model is then constrained by maximizing a Gaussian likelihood function evaluated at a fixed fiducial cosmology, with the free parameters of the covariance matrix as fitting variables. The best-fit covariance is then obtained by ensuring that the $\chi^2$ values from the model match the theoretical distribution for the observed data: the correct covariance should provide  $\chi^2$ values from each simulation that follow a $\chi^2$ distribution.
The key advantage of this approach is that it requires significantly fewer simulations than traditional methods for covariance estimation—typically around $10^2$ simulations, compared to the thousands ($\sim\,10^3-10^4$) usually needed for a full, accurate numerical covariance matrix.

\section{Methodology} \label{sec:methodology}

\begin{table}
	\centering
	\caption{The different cosmological parameters considered in this work.}
	\label{tab:cosm}
	\begin{tabular}{c|c|c|c|c|c}
		\hline
		Name & $\Omega_{\rm m,0}$ & $h$ & $\Omega_{\rm b,0}$ & $n_{\rm s}$ & $\sigma_8$ \\\hline
		$C0$ & $0.3158$ & $0.6732$ & $0.0494$ & $\phantom{-}0.9661$ & $0.8102$ \\
		$C1$ & $0.1986$ & $0.7267$ & $0.0389$ & $\phantom{-}0.9775$ & $0.8590$ \\
		$C2$ & $0.1665$ & $0.7066$ & $0.0417$ & $\phantom{-}0.9461$ & $0.8341$ \\
		$C3$ & $0.3750$ & $0.6177$ & $0.0625$ & $\phantom{-}0.9778$ & $0.7136$ \\
		$C4$ & $0.3673$ & $0.6353$ & $0.0519$ & $\phantom{-}0.9998$ & $0.7121$ \\
		$C5$ & $0.1908$ & $0.6507$ & $0.0527$ & $\phantom{-}0.9908$ & $0.8971$ \\
		$C6$ & $0.2401$ & $0.8087$ & $0.0357$ & $\phantom{-}0.9475$ & $0.8036$ \\
		$C7$ & $0.3020$ & $0.5514$ & $0.0674$ & $\phantom{-}0.9545$ & $0.8163$ \\
		$C8$ & $0.4093$ & $0.7080$ & $0.0446$ & $\phantom{-}0.9791$ & $0.7253$ \\
		\hline
	\end{tabular}
 \tablefoot{The cosmological parameters have been uniformly drawn from the $95$ percent confidence level hyper-volume of the cluster abundance constraints presented in~\citet{DES:2020cbm}.}
\end{table}

We generate halo catalogs using two different approaches: large $N$-body simulations with \og\, and approximate simulations with \pinocchio. Below, we describe the setup of each simulation suite, summarize the resulting halo catalogs used in our analyses, and describe the procedure to measure the 2PCF and power spectra.

\subsection{$N$-body simulations} \label{sec:nbody_sims}

$N$-body simulations are carried out with the Tree-PM \og\ $N$-body code (Dolag et al. in prep.). \og\ represents an evolution of the {\tt GADGET3} code, which in turn provides an improvement of the publicly available {\tt GADGET2} code \citep{Springel:2005mi}. It adopts a mixed MPI/OpenMP parallelization that allows the code to optimally exploit the parallelism offered by computing nodes with a significant amount of shared memory.
Our simulations adopt a flat $\Lambda$CDM cosmology that we refer to as $C0$ throughout this paper and present the specific cosmological parameter values in Table~\ref{tab:cosm}. 

We run 120 realizations of this cosmology, each following $2048^3$ dark matter particles in a comoving, cubic box of side length $3870\,h^{-1}\,\mathrm{Mpc}$. The gravitational softening length is chosen to be one-fortieth of the mean inter-particle spacing, ensuring that the halo mass function and large-scale clustering are well-resolved across the range of halo masses of interest (typically $M \gtrsim 10^{14}\,h^{-1}\,\mathrm{M_\odot}$). The initial conditions (ICs) for 100 of these realizations are generated with the \textsc{Monofonic} code \citep{Michaux:2020yis}. The remaining 20 adopt the same Fourier amplitudes and phases used for a subset of the \pinocchio\ simulations (see Sect.~\ref{sec:pin}) to facilitate a direct comparison.

Each \og\ simulation is evolved from $z=24$ down to $z=0$, outputting snapshots at various intermediate redshifts $z\in\{0.0, 0.5, 1.0, 2.0\}$. We identify halos in each snapshot using \subfind~\citep{Springel:2000qu,Dolag:2008ar,Springel:2020plp}. \subfind\  determines halos by first executing a parallel friends-of-friends with linking length set to $0.2$~\citep[FOF, see,][for instance]{Davis:1985rj} and then assigning its center to the position of the particle with
the lowest potential. The simulation outputs provide a set of \textit{comoving box} halo catalogs for each realization. 

\subsection{Approximate simulations: \pinocchio} \label{sec:pin}

We also use the Lagrangian Perturbation Theory (LPT)-based \pinocchio\ code \citep{Monaco:2001jg,Monaco:2013qta,Munari:2016aut} to produce halo catalogs in a more computationally efficient manner, enabling exploration of multiple cosmological models and a large number of realizations. 

This work complements the original set of \pinocchio\ catalogs presented in~\citet{Euclid:2021api} and serves as a basis for comparison. We generate 20 \pinocchio\ realizations using the $C0$ cosmological parameters and the same mass resolution as our \og\ simulations. These runs employ identical box size, particle number ($2048^3$), and initial Fourier mode amplitudes and phases, enabling a direct one-to-one comparison of halo statistics between \pinocchio\ and full $N$-body catalogs under identical initial conditions. In contrast, the original set assumed $\Omega_{\rm m,0}=0.30711$ and used a grid of $2160^3$ elements, which led to a mass resolution difference of roughly 20 percent.

We further exploit \pinocchio\ efficiency to create an extensive suite of 900 comoving box simulations covering nine additional cosmological models, labeled originally $C0$ through C8 in \citet{Euclid:2022dbc}. Each cosmology has 100 realizations, each with $3072^3$ dark matter particles in a box the same size as the previously presented simulations. These simulations broadly sample parameter space, enabling us to study how our covariance model (and its nuisance parameters) might vary across different cosmological parameters.

In addition to comoving outputs, \pinocchio\ can generate halo catalogs on past light cones. We construct such cones with a $60\deg$ aperture. These light-cone catalogs mimic in a simplistic way the typical sky coverage of future wide-field cluster survey footprints on an idealized uniform and unmasked coverage.

\subsection{Summary of simulation sets}
Overall, we have assembled:
\begin{enumerate}
    \item 900 high-resolution \pinocchio\ realizations spanning nine distinct cosmological models ($C0$--$C8$), each with $3072^3$ particles and box size of $3870\,h^{-1}\,\mathrm{Mpc}$, for robust exploration of cosmology dependence. We further generated catalogs in the past light cone of 60\,$\deg$ aperture for these simulations, mimicking wide-field cluster surveys;
    \item 100 low-resolution $N$-body realizations at the $C0$ cosmology with the same box size and $2048^3$ particles; 
    \item 20 \pinocchio\ and 20 $N$-body realizations were generated at low resolution ($2048^3$ particles) assuming the $C0$ cosmology. Both simulations employed identical phases and amplitudes of the initial Fourier modes, thereby allowing straightforward comparisons of halo statistics within comoving boxes of identical size. Because the \pinocchio\ and \og\ runs in this sub-set share the same initial conditions, the sample noise in the relative statistics between the two codes is greatly reduced;
    \item 100 \pinocchio\ realizations from the set used in~\citet{Euclid:2021api,Euclid:2022txd}, with low resolution similar to the $N$-body set. These mocks, labeled as ``original'' set, have the same box size and a cosmology similar to $C0$. This last set will serve as a reference for comparison with previous works. 
\end{enumerate}

The combination of full $N$-body and approximate \pinocchio\ simulations at matched initial conditions and over multiple cosmologies allows us to assess the robustness of the 2PCF covariance model, as well as the cosmology dependence of the nuisance parameters introduced in Sect.~\ref{sec:theory}.

\subsection{Post-processing} \label{sec:postproc}

\subsubsection{Mass rescaling}

\pinocchio\ has been calibrated to reproduce within a few percent the FOF-based HMF presented by~\citet{Watson:2012mt}. To reproduce instead our adopted model for the HMF~\citep{Euclid:2022dbc} while keeping the sample variance and shot-noise for each simulation, we implement the rescaling of the halo masses as proposed by~\citet{Euclid:2021api} on all \pinocchio\ outputs. In order to obtain a fair comparison in Sect.~\ref{sec:resolution}, we also rescale the original set of \pinocchio\ mocks to the~\citet{Euclid:2022dbc} HMF. The same procedure was applied to the \og\ halo catalogs to keep consistency between the catalogs. Given the proven accuracy of our fiducial HMF calibration presented in~\citet{Euclid:2022dbc}, the impact of the recalibration for the $N$-body simulation is rather small.

\subsubsection{Power-spectra}

We used the set of PYthon LIbraries for the Analysis of Numerical Simulations (\pylians)\footnote{\url{https://github.com/franciscovillaescusa/Pylians}} to construct the density field and compute the power spectra from the halo catalogs. For the 20 simulations that share the initial conditions between \og\ and \pinocchio, the cross-spectra between matter and halos are calculated using the linear density field produced by \pinocchio. All power spectra were calculated from a 3D grid with $512^3$ mesh points on which the density traced by the halo distribution is assigned with a Cloud In Cell (CIC) scheme. The power-spectra measurements are averaged within shells in $k$-space with widths given by $k_f\equiv 2\pi/L$, corresponding to the fundamental mode of the box.

\subsubsection{2PCF}
We measure the 2PCF by comparing halo pair distributions in the data and random catalogs using the estimator from \citet{Landy:1993yu}. The random catalog is created by extracting and shuffling positions of a subset of halos from each mock catalog, generating a catalog with \(n_{\rm R} = 20\,n_{\rm D}\) objects, randomly distributed within the box/lightcone volume.

The 2PCF is computed for halos with masses above $10^{14}\,h^{-1}\,M_\odot$ and $3 \times 10^{14}\,h^{-1}\,M_\odot$, in 25 log-spaced bins in the separation range $r = 20$\,--\,$130\,h^{-1}$\,Mpc, which includes linear scales, where the bias is almost constant, and the BAO peak. As for the analysis of the lightcones, we consider four redshift bins of width \(\Delta z = 0.5\) in the range $z = 0$\,--\,$2$. The pair counts in the data-data, data-random, and random-random catalogs are measured using the \texttt{CosmoBolognaLib} package \citep{Marulli:2015jil}.

\section{\label{sec:results}Results}

This section provides a systematic comparison of the 2PCF and its covariance measured from the simulation sets introduced in Sect.~\ref{sec:methodology}. We assess resolution effects in \pinocchio\ (Sect.~\ref{sec:resolution}), comparing C0 simulations from ``set 1'' to original simulations in ``set 4'', as well as \pinocchio\ simulations in ``set 3''. Next, we directly compare \pinocchio\ and \og\ simulations with identical initial conditions (``set 3'', Sect.~\ref{sec:pinocchio_gadget}), quantifying the agreement in bias and shot-noise corrections, as well as in the covariance parameters fitted from ``set 1'' and ``set 2''. Finally, we examine the cosmology dependence of the covariance-model parameters (Sect.~\ref{sec:cosmo}), using all the ``set 1'' simulations.

\subsection{\label{sec:resolution}Resolution}

\begin{figure*}[h]
\centering
\includegraphics[width = \textwidth]{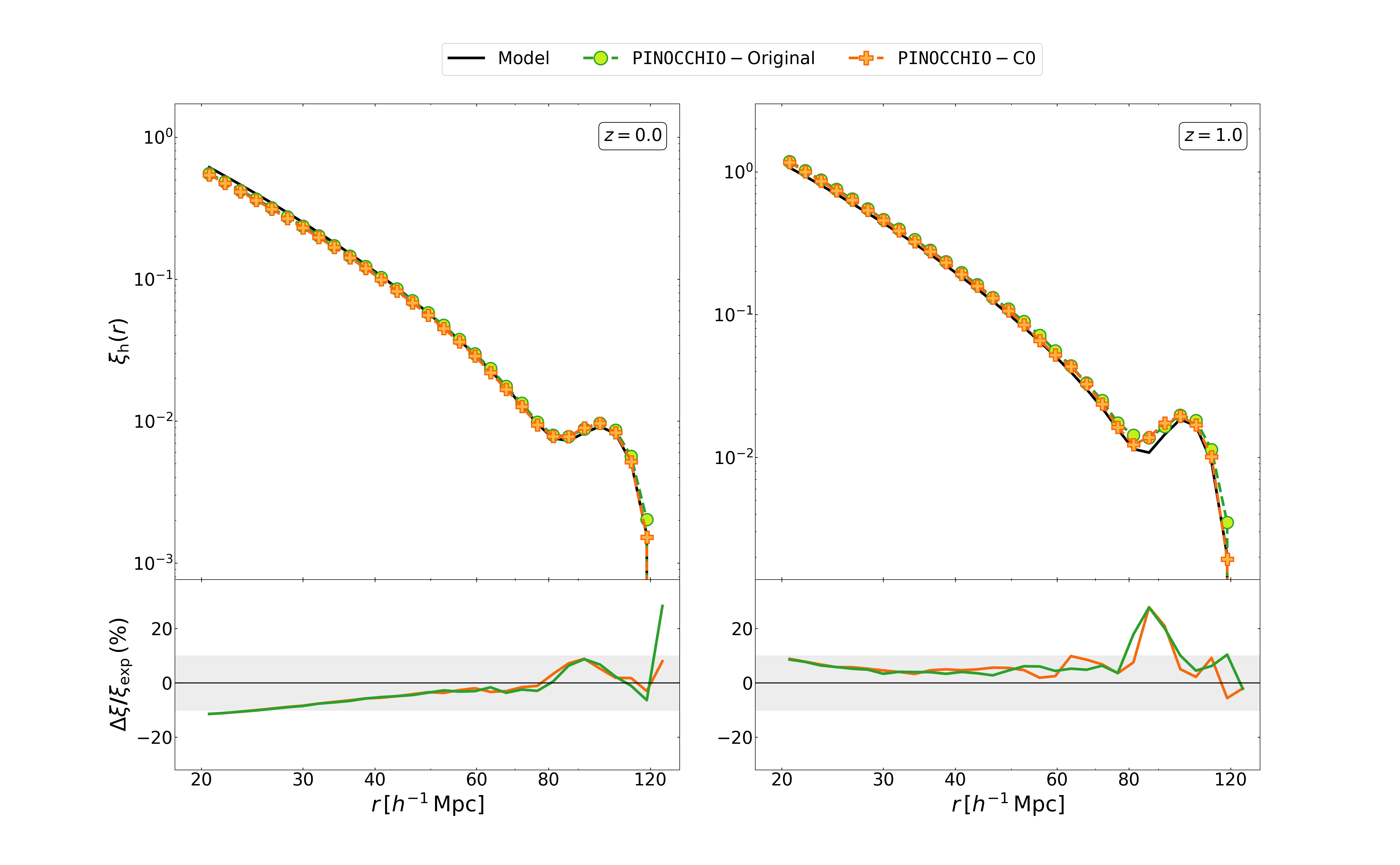}
\caption{\emph{Top:} Comparison of the \pinocchio\ 2PCF measured on the comoving box from the original mocks~\citep{Euclid:2021api} and our new high-resolution $C0$. The new and original simulations assume different but similar cosmologies. We present only the theoretical expectation for the $C0$ cosmology for better figure readability. \emph{Bottom:} residual between the measured 2PCF and the respective theoretical expectation. The shaded gray area denotes the region within a relative difference smaller than 10 per cent.}
\label{fig:2pcf_pin}
\end{figure*}

\begin{figure*}[h]
\centering
\includegraphics[width = \textwidth]{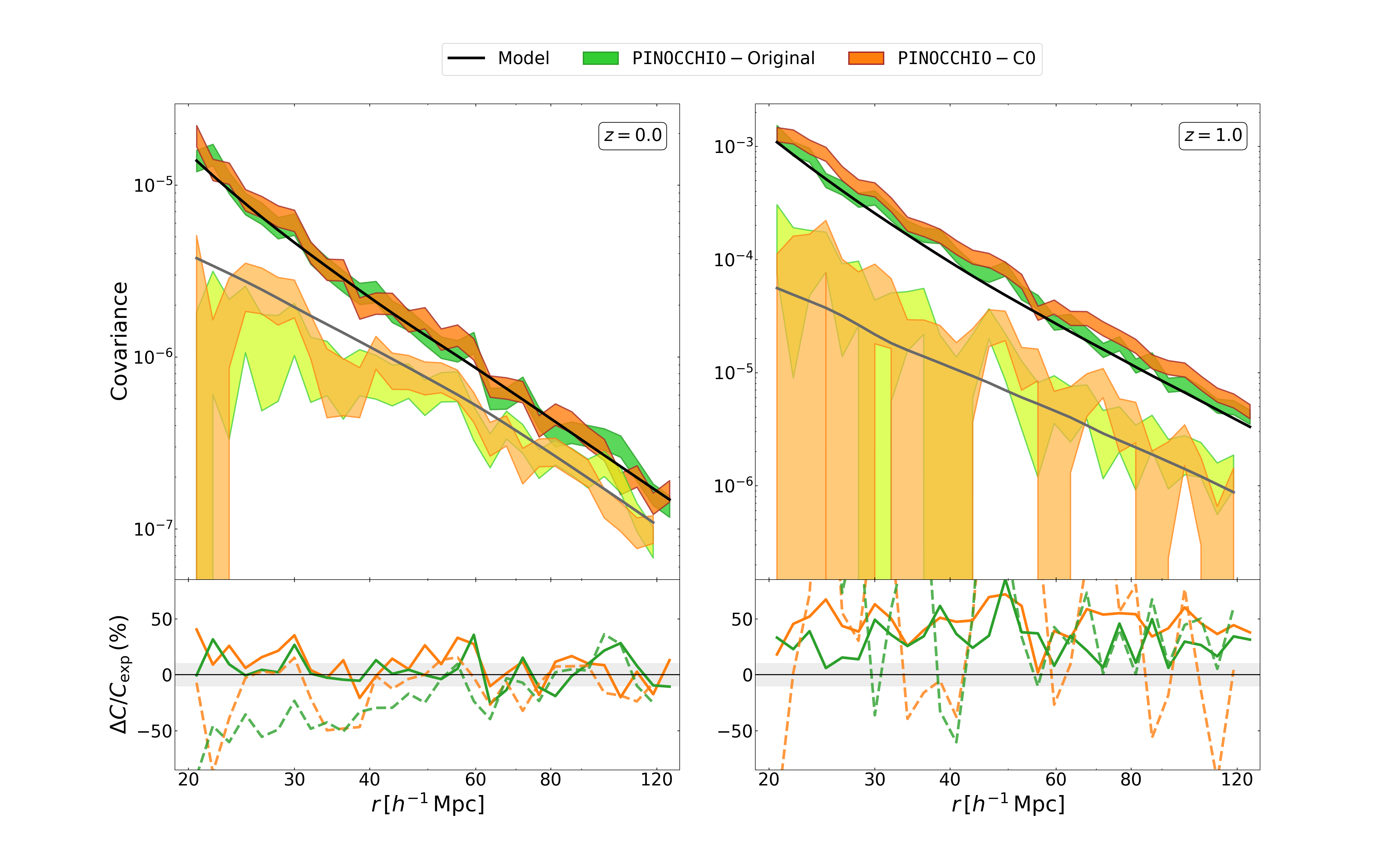}
\caption{\emph{Top:} Comparison of the \pinocchio\ covariance of the 2PCF measured on the comoving box from the original mocks~\citep{Euclid:2021api} and our new high-resolution $C0$. Similarly to Fig.~\ref{fig:2pcf_pin}, we present only the theoretical expectation from the baseline model presented in Eq.~\eqref{eq:cov_baseline} for the $C0$ cosmology for better figure readability. \emph{Bottom:}  residual between the measured 2PCF and the respective theoretical expectation. In both panels, darker colors are used for diagonal and lighter for the first off-diagonal terms.}
\label{fig:cov_pin}
\end{figure*}

\begin{figure*}[h]
\centering
\includegraphics[width = \textwidth]{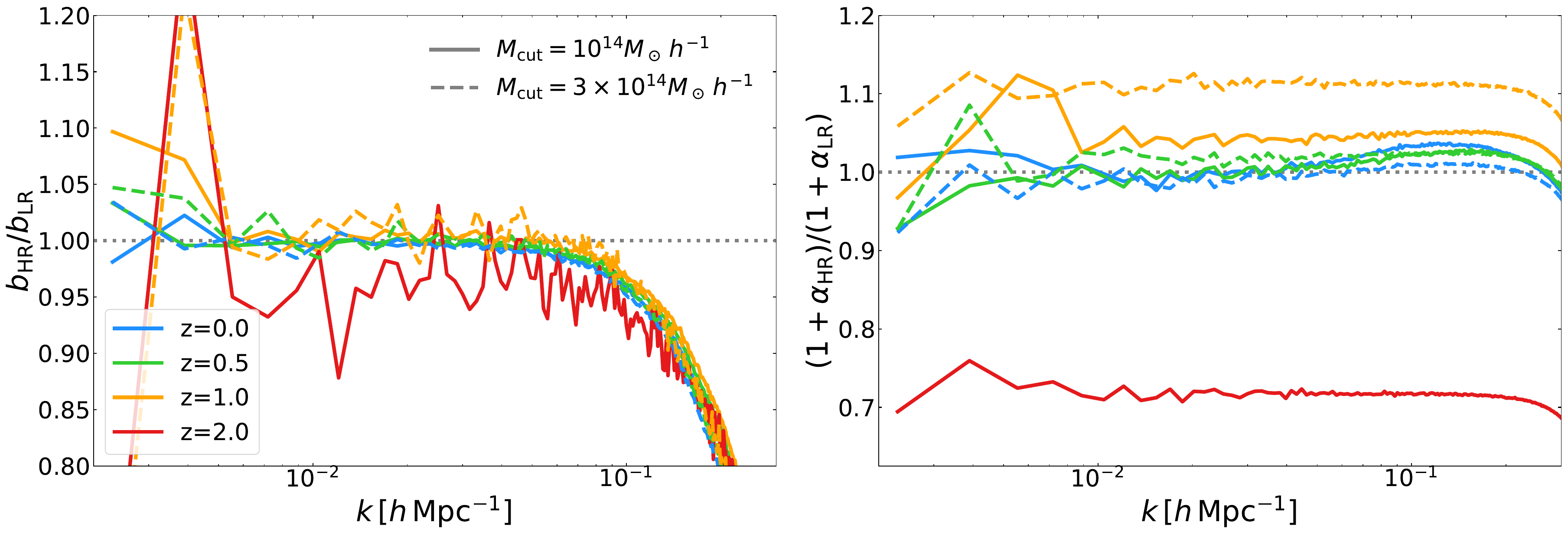}
\caption{Comparison of resolution effects on \pinocchio\ predictions for halo bias $b(k)$ (left) and shot-noise correction $\alpha(k)$ (right). We show the ratio between high-resolution (HR: 100 $C0$ simulations) and low-resolution (LR: 20 simulations) results at redshifts $z\in\{0.0,0.5,1.0,2.0\}$ for two mass thresholds: $M_{\rm cut}=10^{14}\msun$ (solid) and $M_{\rm cut}=3\times10^{14}\msun$ (dashed). While bias differences remain below two percent for $z\leq1.0$, reaching 5 percent at $z=2$ (the larger mass cut case is omitted for $z=2$  due to noise), the shot-noise corrections converge within 10 percent for $z\leq1$ but show 30 percent deviations at $z=2$.}
\label{fig:pin_res}
\end{figure*}

In Fig.~\ref{fig:2pcf_pin}, we present a comparison between the 2PCF measurement at redshifts $z=0$ and $z = 1$ for the high-resolution $C0$ \pinocchio\ simulations (``set 1'') and the low-resolution original set (``set 4''). The cosmological parameters from the original simulations and the $C0$ ones differ. Still, the difference in the theoretical expectation is too small to be seen on the logarithmic scale, and therefore, we present it only for the $C0$ case. In the bottom panel, we present the residual between the measurements and the corresponding theoretical measurements. We observe similar trends between the simulation sets at both redshifts. At redshift $z=0.0$, \pinocchio\ reproduces the theoretical expectation within 10 percent for the range of separation $r \in (35-115)\,\mpc$. The agreement between \pinocchio\ and expectations generally improves for the case $z=1.0$. Still, we notice that in the region $r \sim 85\,\mpc$ the agreement worsens. This fluctuation is due to the binning where the steepness of the 2PCF close to the BAO peak causes an artificially large impact. In the residual panel of Fig.~\ref{fig:2pcf_pin}, we observe that the new and original sets present similar behaviors, ruling out possible resolution effects from the original calibration. 

Similarly to Fig.~\ref{fig:2pcf_pin}, in Fig.~\ref{fig:cov_pin}, we compare the baseline 2PCF covariance presented in Eq.~\eqref{eq:cov_baseline} between the original set and the $C0$ subset. We present the results for the diagonal (darker colors) and first off-diagonal terms (lighter colors). The similar behavior between the original and new simulations, which have higher resolution, confirms the claim of~\cite {Euclid:2022txd} that the baseline model cannot account for the observed covariance in simulations without calibration of the parameters $\alpha$, $\beta$, and $\gamma$.

To assess the impact of the resolution on the \pinocchio\ predictions, in Fig.~\ref{fig:pin_res}, we compare the bias $b$ and shot-noise correction $\alpha$ measured from the new 100 $C0$ simulations at higher resolution with respect to the $20$ simulations with same cosmology but lower resolution. The measurements are from the outputs in comoving boxes at different redshifts and with different minimum mass selection $M_{\rm cut} \in \{10^{14}, 3\times10^{14}\}\,\msun$. We measured the bias from the ratio of the cross-spectrum $P_{\rm h,m}$ between the halo and matter to the matter-power spectrum $P_{\rm m}$. The shot-noise correction $\alpha$ is measured using the following relation
\begin{equation}
    \alpha(k)+1 = \bar{n}(z)\,\left( P_{\rm h}(k, z) - \frac{P_{\rm h,m}(k, z)^2}{P_{\rm m}(k, z)}\right)\,,
    \label{eq:alpha}
\end{equation}
where $\bar{n}(z)$ is the mean density of halos.

We observe in Fig.~\ref{fig:pin_res} that the differences in the bias due to resolution are more minor than $2$ percent for the different mass and redshift selections, except the redshift two, where differences are roughly 5 percent for $10^{14} \msun$ mass selection. The case for the mass selection of $3\times10^{14} \msun$ is not shown for redshift two as, in this case, the measurements are very noisy due to the small number of tracers.
Concerning the shot-noise correction $\alpha$, we observe that the original simulations converge within a few percent for redshifts below unity. For redshift $z=1$, while the measurements from the $10^{14}\msun$ still agree within said accuracy, for the $3\times10^{14}\msun$, we observe a difference of the order of $10$ percent. The differences can be as significant as $30$ percent for the case of redshift $z=2$. This indicates that the original resolution might be too low to calibrate the shot-noise correction accurately at high redshift.

Notice that the parameters $\alpha$
and $\beta$ in the covariance model have slightly different meanings and values compared to those fitted from the power spectrum. This is because the three covariance parameters also absorb the effects of missing higher-order terms in the covariance, partially losing their physical interpretation~\citep{Euclid:2022txd}. The agreement between these parameters will be discussed in Sect.~\ref{sec:pinocchio_gadget}.

\subsection{\label{sec:pinocchio_gadget}\og\ vs \pinocchio\ }

\begin{figure*}
    \centering
    \includegraphics[width=\textwidth]{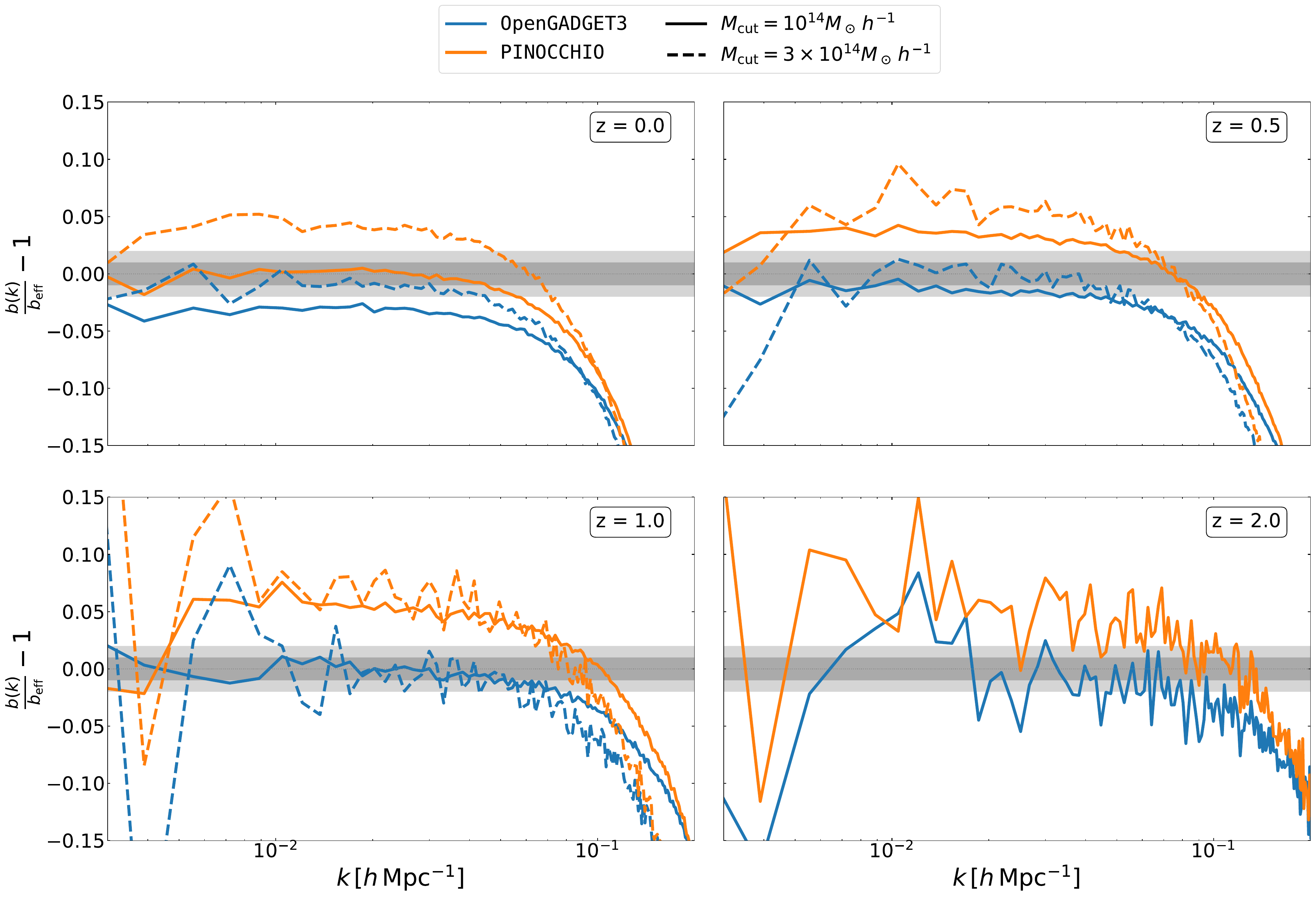}
    \caption{Comparison between the effective bias from mass-scaled halo catalogs from \og\ and \pinocchio\ and the theoretical expectation from~\citet{Euclid:2024wog}. Different panels show the results for different redshifts corresponding to $\{0.0, 0.5, 1.0, 2.0\}$ clockwise. Different line styles correspond to different minimum mass selections. We omit the result for the $M_{\rm cut} = 3\times10^{14}\, h^{-1}M_\odot$ at $z=2$ as the measurements are too noisy due to the low statistics.}
    \label{fig:bias_comp}
\end{figure*}

\begin{figure*}
    \centering
    \includegraphics[width=\textwidth]{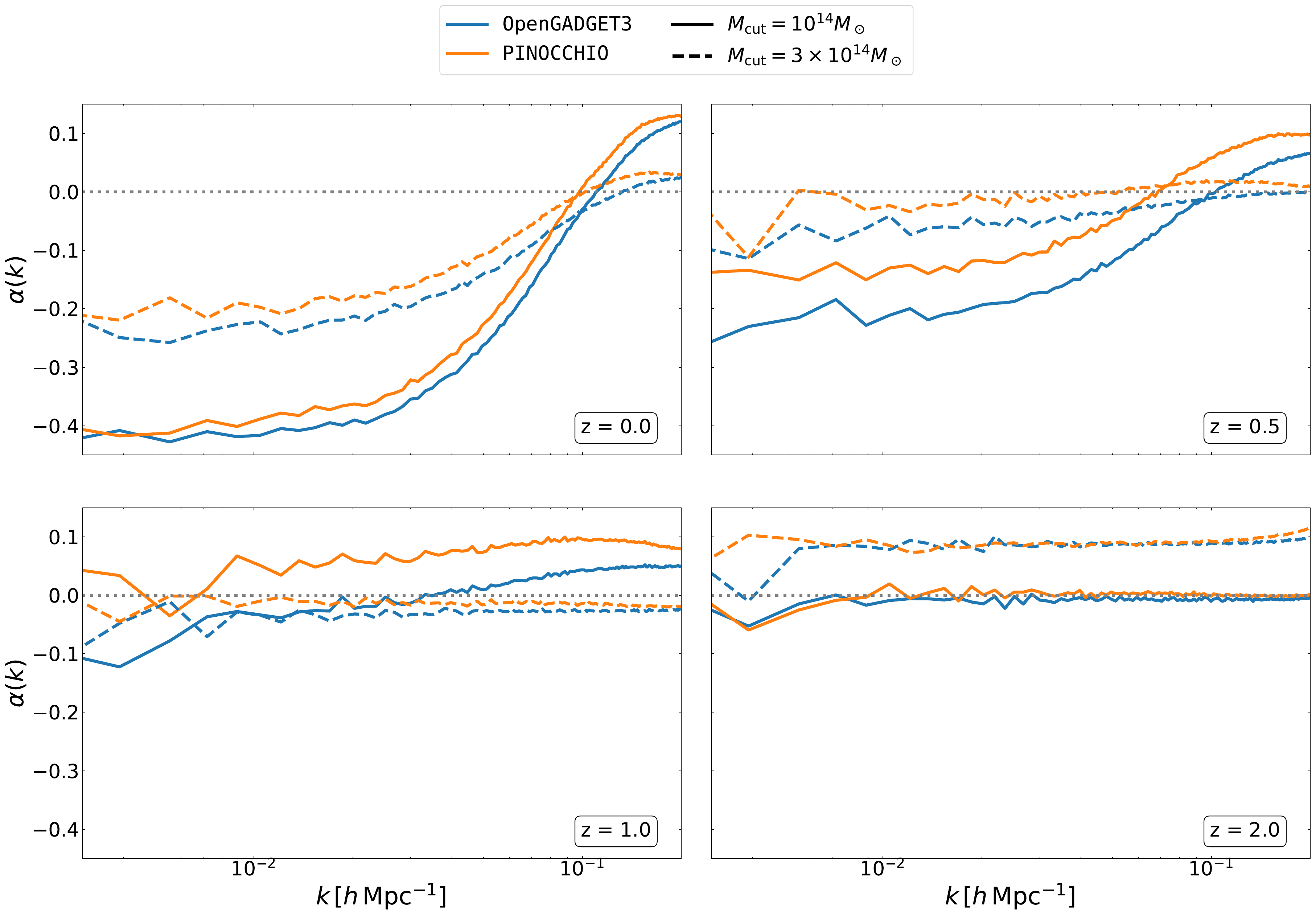}
    \caption{Comparison between the non-Poissonian correction to the shot-noise of the mass-scaled halo catalogs from \og\ and \pinocchio. Different panels show the results for different redshifts corresponding to $\{0.0, 0.5, 1.0, 2.0\}$ clockwise. Different lifestyles correspond to different minimum mass selections.}
    \label{fig:alpha_comp}
\end{figure*}

In Fig.~\ref{fig:bias_comp}, we compare the effective halo bias measured from the mass-scaled catalogs produced by \og\ and \pinocchio\ (``set 3'') to the theoretical prediction from \citet{Euclid:2024wog}. 
Overall, \og\ and \pinocchio\ agree at the 5--10\% level for most scales and redshifts, but at higher 
redshifts (particularly $z=2.0$) and for larger mass cuts, we observe slightly larger deviations. These differences between the codes reflect residual non-linearities or shot-noise corrections not fully captured by the mass-scaling procedure. 

In Fig.~\ref{fig:alpha_comp}, we compare the non-Poissonian correction to the shot noise, $\alpha(k)$, for the same halo catalogs. 
We find that both \og\ and \pinocchio\ exhibit broadly similar trends in $\alpha(k)$, remaining close to zero for large scales and high redshifts but displaying noticeable deviations for high masses and lower redshifts. 

Figures~\ref{fig:bias_comp} and \ref{fig:alpha_comp} highlight that, when using the same initial phases and a consistent mass-scaling scheme, \og\ and \pinocchio\ reproduce the large-scale bias and shot-noise properties of halo catalogs to within a few percent in most regimes. These results suggest that our mass-scaling procedure effectively aligns the mass functions between \og\ and \pinocchio. Still, subtle discrepancies remain, likely reflecting that the scaling procedure does not consider each cluster's environment.

\begin{figure*}[h]
\centering
\includegraphics[width=0.49\textwidth]{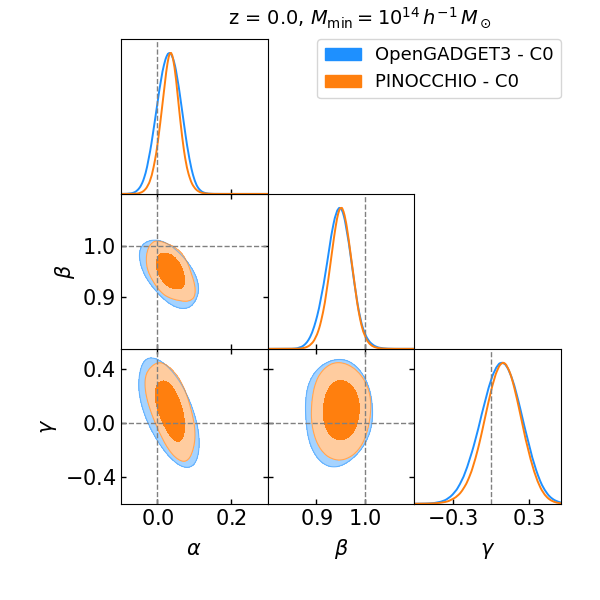}
\includegraphics[width=0.49\textwidth]{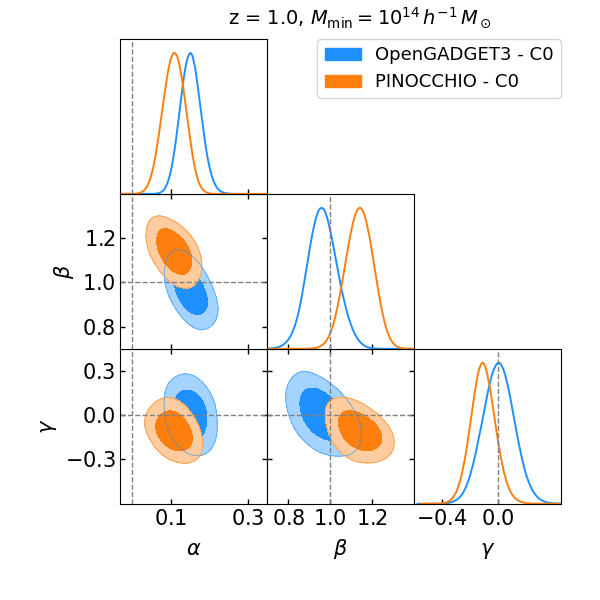}
\includegraphics[width=0.49\textwidth]{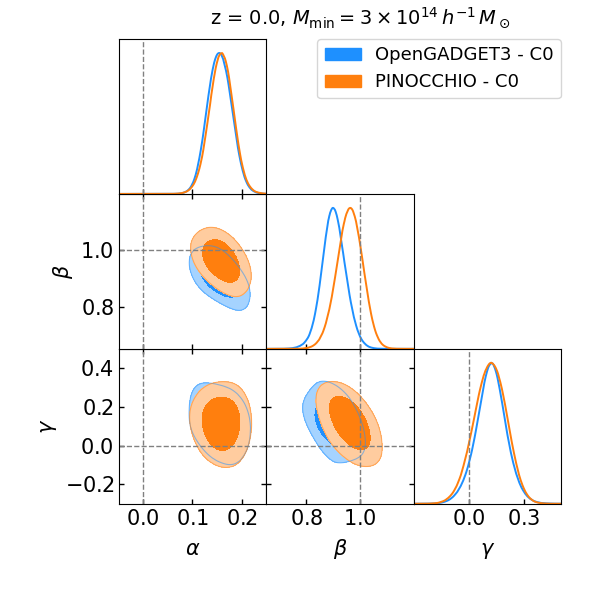}
\includegraphics[width=0.49\textwidth]{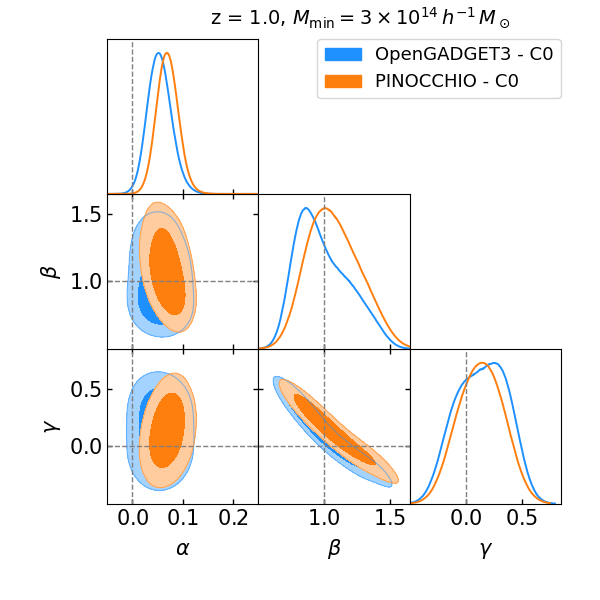}
\caption{Posteriors for fitted covariance parameters: \pinocchio\ vs \og\, for the C0 cosmology. On the \textit{left panels} redshift $z=0$, on the \textit{right panels} redshift $z=1$. \textit{Top panels:} $M > 10^{14} \,h^{-1} M_\odot$, \textit{Bottom panels:}$M > 3\times10^{14}\,h^{-1}\, M_\odot$.}
\label{fig:fit_box}
\end{figure*}

\begin{figure*}[h]
\centering
\includegraphics[width = \textwidth]{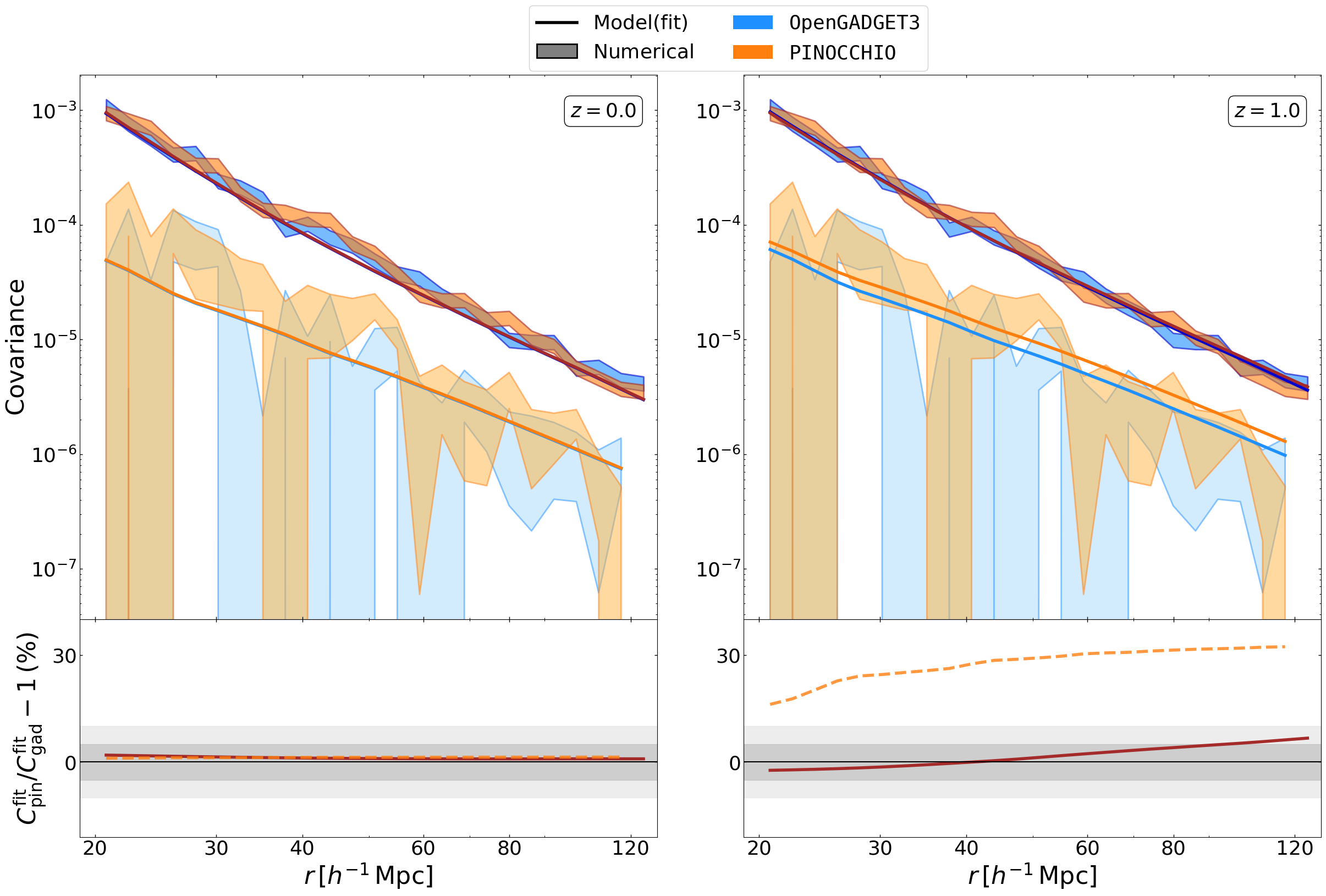}
\caption{Covariance comparison (box): \pinocchio\ vs \og\ covariance matrices, at $z=0$ (left panel) and $z=1$ (right panel). Shaded areas represent the numerical covariance within 1$\sigma$ uncertainty, and solid lines are the semi-analytical prediction, with fitted parameters. Orange colors indicate \pinocchio\ covariance, while blue colors are for \og\; darker colors represent the diagonal terms, lighter colors are the first off-diagonal terms. Bottom panels show the residuals of the \pinocchio\ fitted model with respect to the \og\ one (solid line for the diagonal elements, dashed line for the off-diagonal ones).}
\label{fig:2cov_gad}
\end{figure*}

Figure~\ref{fig:fit_box} presents the posterior distributions of the parameters  $\alpha$, $\beta$, and $\gamma$ from Eq.~\eqref{eq:cov_fit}, derived by fitting the 2PCF covariance measured in the comoving boxes for \og\ (``set 2'') and \pinocchio\ (``set 1''). 

Overall, we observe good agreement between the two codes for both mass cuts. The values of $\alpha$ and $\gamma$ are broadly consistent across the different simulations, reflecting similar levels of non-Poissonian noise corrections and subleading shot-noise contributions. A slight offset emerges in $\beta$, with \pinocchio\ typically favoring higher values than \og, in line with the differences in the practical bias seen in Fig.~\ref{fig:bias_comp}. This offset suggests that the \pinocchio\ halos may require a larger rescaling of the bias to reproduce the measured covariance, possibly reflecting the limitations of the approximated method. Nonetheless, the overlap in the posterior contours for all parameters indicates that the two approaches yield broadly consistent covariance calibrations.
Evidence for this can be observed in Fig.~\ref{fig:2cov_gad}, which compares the numerical and semi-analytical covariances for \og\ and \pinocchio, using best-fit parameters from Fig.~\ref{fig:fit_box}. The residuals between the two models show agreement within 5\% in most cases, except for off-diagonal terms at $z=1$. However, these terms are subdominant and do not significantly impact cosmological analysis.
Such a result reinforces the conclusion that \pinocchio\ can reliably match \og\ covariance results with an appropriate mass-scaling procedure and modest adjustments of the nuisance parameters.

We also find that fitting the covariance parameters using 100 mocks yields stable results up to 25 separation bins within our chosen radial range. Although this is not a universal criterion, more than 100 simulations may be required if a finer binning scheme is used.

\subsection{\label{sec:cosmo}Cosmological dependency}

\begin{figure}
    \centering
    \includegraphics[width=0.5\textwidth]{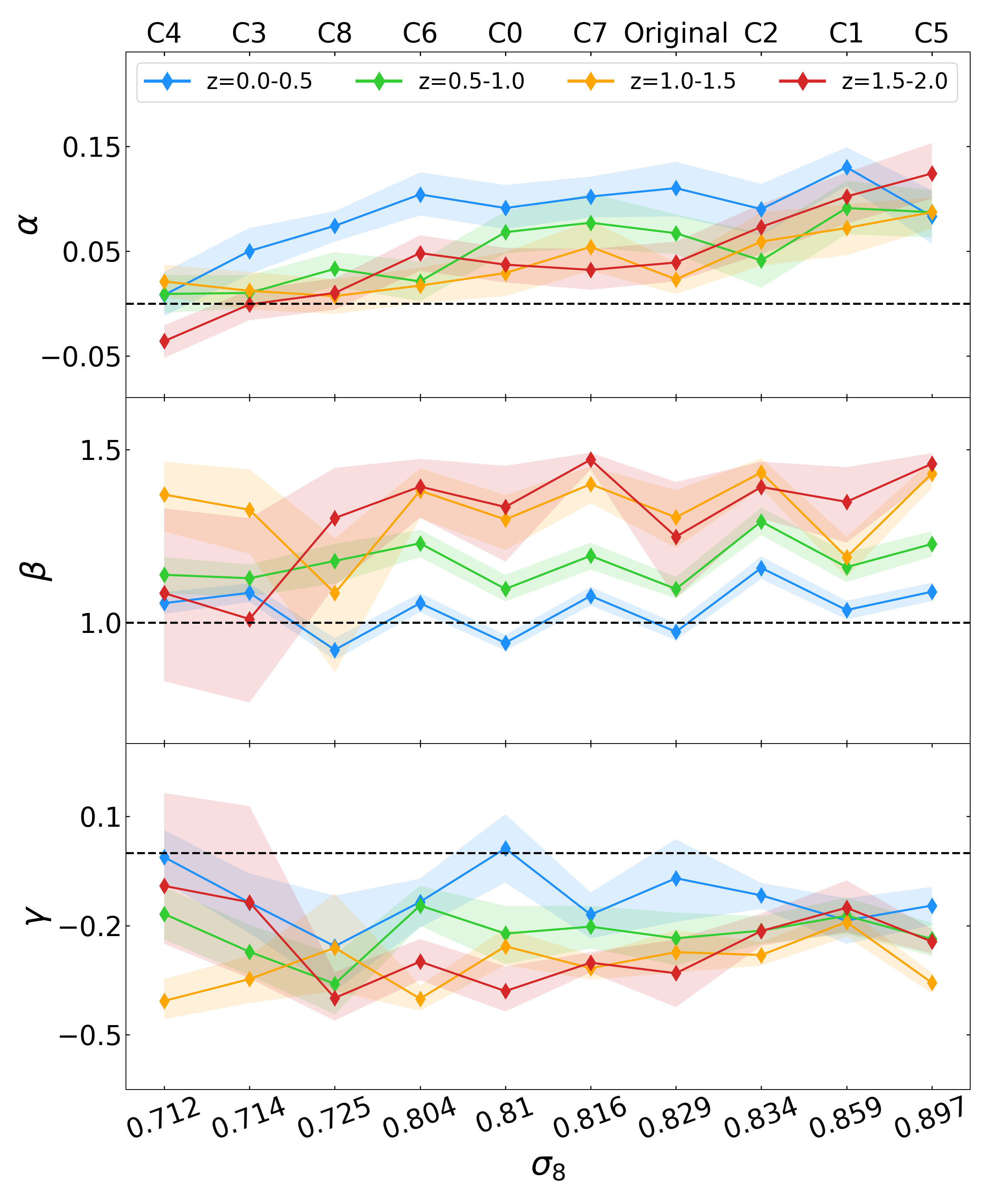}
    \caption{Best-fit covariance-model parameters \(\alpha\) (top panel), \(\beta\) (middle panel), and \(\gamma\) (bottom panel) as a function of cosmology. 
    Each color corresponds to a different redshift bin 
    \(\left(z=0.0\text{--}0.5,\,0.5\text{--}1.0,\,1.0\text{--}1.5,\,1.5\text{--}2.0\right)\) 
    in the light cone, with shaded bands indicating the statistical uncertainties. 
    The dashed black lines mark the reference values \(\alpha=0\), \(\beta=1\), and \(\gamma=0\). 
    The cosmologies on the upper x-axis are sorted by increasing values of \(\sigma_8\). }
    \label{fig:cov_cosmo_fit}
\end{figure}

In Fig.\ref{fig:cov_cosmo_fit}, we show the best-fit values for the covariance model from Eq.\eqref{eq:cov_fit} as a function of cosmology. For a more realistic comparison, we perform the fit on lightcones divided into different redshift bins. The parameter $\alpha$, while staying within the range 
$[-0.05, 0.15]$—indicating small deviations from Poissonian shot noise—tends to increase with $\sigma_8$, suggesting a mild linear dependence on the amplitude of matter fluctuations. By contrast, $\beta$ (the bias rescaling) and $\gamma$ (the sub-leading shot-noise correction) do not show a strong trend with $\sigma_8$. However, both $\beta$ and $\gamma$ exhibit noticeable variations with redshift. Specifically, $\beta$ remains in the range $[1.0,\,1.5]$ for the different cosmologies but shifts slightly across redshift bins. At the same time, $\gamma$ is negative for most cosmologies, implying that the baseline Poisson noise may be an overestimate of the actual sampling noise in specific regimes but tends to decrease with increasing redshift. These results emphasize that even though the global amplitude of non-Poissonian effects and bias corrections is small, allowing for flexible parameters in the covariance model is essential to capture the redshift dependencies and subtle but non-trivial responses to variations in $\sigma_8$.

Additionally, the amplitude of $\beta$ typically exceeds the observed accuracy of \pinocchio\ in reproducing the linear bias, suggesting that $\beta$ may absorb extra flexibility beyond merely correcting the bias (e.g., mild non-linearities). Furthermore, the redshift dependence of both $\beta$ and $\gamma$ indicates that these parameters are sensitive to the survey selection function. Consequently, when applying the model to cosmological data, one should calibrate these parameters on realistic mocks that replicate the actual selection function, ensuring that the resulting covariance matrix accurately captures both the intrinsic clustering of the sample and the observational selection effects.

Although we show the results for increasing $\sigma_8$, we have also analyzed the residual dependency as a function of $\Omega_{\rm m}$ and the composite variable $S_8=\sigma_8\,\sqrt{\Omega_{\rm m}/0.3}$. While $\alpha$ shows some small dependence on $S_8$, it is less significant than the dependence on $\sigma_8$; the other parameters do not show any considerable dependency. 

The dependency of the non-Poissonian correction primarily on $\sigma_8$ and not on $S_8$ is indicative that the clustering amplitude of the underlying density field is the physical mechanism behind it and not the number density of halos of a given mass, which is instead determined by $S_8$.
Non-linear clustering indeed introduces correlations between the fluctuations in the halo number density that could be reflected in the deviation of the density field sampling provided by halos from a standard Poisson sampling. A thorough study of this issue deserves a dedicated analysis that goes beyond the scope of this work.

\section{\label{sec:discussion}Discussion}
Approximate methods, and in the specific case of this paper, the \pinocchio\ code, have become a valuable tool in cosmology by providing results in a fraction of the time required by full $N$-body simulations~\citep{Chuang:2014toa,Monaco:2016pys}. However, this efficiency is achieved by compromising the accuracy with which the underlying statistical properties are reproduced. Different papers have shown that \pinocchio\ nominal accuracy for different statistics, including subhalo clustering data~\citep{Berner:2021hmp} and full-sky 21\,cm intensity mapping mocks~\citep{Hitz:2024cwl}, is roughly 10 percent. Due to its efficiency, \pinocchio\ is a natural candidate to generate massive synthetic data needed to extract the covariance of different cluster statistics numerically~\citep{Euclid:2021api,Euclid:2022txd}. However, as shown by~\citet{Sellentin:2019ehy}, approximate covariance matrices might bias cosmological constraints and, therefore, should be treated at equal footing with the signal modeling to ensure an unbiased analysis.

In this paper, we have focused on the cluster clustering covariance. In~\citet{Euclid:2022txd}, it was shown that the low-order analytical model of~\citet{Meiksin:1998mu} does not quantitatively reproduce the numerical covariance extracted from~\pinocchio\ catalogs. \citet{Euclid:2022txd} proposed a semi-analytical extension, presented in Sect.~\ref{sec:theory}. The proposed model could reproduce the covariance accurately if the free parameters $\alpha$, $\beta$, and $\gamma$ appearing in Eq.\ref{eq:cov_fit} were appropriately calibrated~\citep{Fumagalli:2022plg}. Therefore, it is important to assess the resolution, cosmological, and methodological impacts on the calibration of these parameters.

The parameters $\alpha$ and $\beta$ are related to the shot-noise correction to the Poisson expectation and residual errors on the linear bias. The non-Poissonian correction can be measured from the simulations using the matter, halo, and matter-halo power spectrum using the estimator presented in Eq.~\eqref{eq:alpha}. The need for a bias correction can be assessed by comparing the halo bias, as measured from the ratio of the matter-halo cross-spectrum and matter power-spectrum at linear scales, concerning the fiducial linear bias model~\citet{Euclid:2024wog}. In Sect.~\ref{sec:resolution}, we show that a mass resolution of approximately $6\times10^{11}\,h^{-1}M_\odot$ for \pinocchio\ is too low to produce convergent results for $\alpha$ and $\beta$, especially at higher redshifts, compared to a simulation that has a factor of 4 more resolution elements. 

Regarding the comparison between \pinocchio\ and \og, in Sect~\ref{sec:pinocchio_gadget}, we show that the shot-noise correction displays similar trends in both codes. However, the overall amplitude of the shot-noise correction (i.e., $1+\alpha$) differs by as much as 10 percent. In contrast, for the bias correction $\beta$, \pinocchio\ systematically overpredicts the value relative to \og, with a stable relative difference of approximately 5 percent that could be easily corrected during calibration. Correcting this slight bias would yield an even better agreement between the calibrations of \pinocchio\ and \og. Nevertheless, even without such a correction, the covariances obtained from both codes remain statistically equivalent within the uncertainties assessed in this work. 

Furthermore, regarding the cosmological dependency of the parameters, only $\alpha$ shows a small dependency with $\sigma_8$. While the linear coefficients of said relation do not show a redshift dependency, its normalization decreases with redshift over the redshift range $z=$[0--1] and is constant at larger redshift.

Lastly, for both \pinocchio\ and \og\, we observe that the extra parameters $\alpha$, $\beta$, and $\gamma$ depend on redshift and mass cuts. This dependency implies that the final covariance depends on the selection function. Therefore, future cosmological analysis should use mock catalogs that mimic the sample selection function.

\section{\label{sec:conclusions}Conclusions}
In this paper, we have assessed the robustness of the calibration of the semi-analytical model for the 2PCF covariance on both fast approximate \pinocchio\ and \og\ $N$-body simulations across multiple cosmological models. We summarize our main findings:

\begin{itemize}
    \item The shot-noise corrections and bias calibration converge more reliably in higher-resolution \pinocchio\ simulations ($M_{\rm part.}\lesssim10^{11}\,h^{-1}M_\odot$), particularly at $z\gtrsim1$. We observe that insufficient mass resolution can bias the non-Poissonian shot-noise parameter by up to 30\%.
    
    \item When started from the same initial random field, \pinocchio\ and \og\ yield broadly consistent covariance estimates for the cluster 2PCF. Small discrepancies at the 5--10\% level can be accounted for by modest shifts in the model’s nuisance parameters (see Eq.~\ref{eq:cov_fit}). This demonstrates that \pinocchio\ can provide a reliable and computationally efficient alternative  
    to full $N$-body simulations to measure clustering covariances for large cosmological surveys of galaxy clusters.
    
    \item Of the three calibrated nuisance parameters, only the non-Poissonian shot-noise correction $\alpha$ shows a mild linear dependence on $\sigma_8$. The other parameters, $\beta$ (bias rescaling) and $\gamma$ (subleading shot-noise correction), exhibit more pronounced redshift and mass cut dependence than cosmology dependence.
    
    \item The redshift and mass cuts and redshift dependencies indicate that the survey selection function impacts the covariance calibration. Accurate modeling of the covariance for application to specific surveys
    thus requires realistic mocks replicating the observed selection to avoid 
    systematic biases in cosmological inference.
\end{itemize}

Overall, our results confirm that this semi-analytical covariance model can robustly capture the main sources of uncertainty in cluster clustering, provided the nuisance parameters are carefully tuned. In particular, approximate simulations such as \pinocchio\ can be employed effectively for such a calibration, enabling robust covariance estimates across diverse cosmological models and redshifts without incurring the computational expense of full $N$-body simulations. Furthermore, the demonstrated success of \pinocchio\ in reproducing the cluster covariance motivates the extension of our analysis to a multi-probe context in future studies. Importantly, assessing its capability to estimate covariances for additional cluster-based summary statistics, such as cluster shear correlation and cluster galaxy correlation, will allow a more comprehensive evaluation of approximate methods in various cosmological applications.

\section*{Data availability}
The numerical data generated and analyzed in this study are not publicly available due 
to the large storage and computational resources required. The authors will 
make every reasonable effort to provide access to the underlying simulation outputs 
and derived data products upon request, subject to resource availability. The fiducial models for the HMF and halo bias used in this paper can be accessed in~\citet{Castro_CCToolkit_A_Python_2024}.\footnote{\url{https://github.com/TiagoBsCastro/CCToolkit}}

\begin{acknowledgements}
The authors A. Fumagalli and T. Castro share co-authorship of this paper. T. Castro conceptualized and produced the simulated data, while A. Fumagalli carried out the formal analysis of the simulated data. All authors contributed to interpreting the results and writing the manuscript. All authors have read and approved the final version of the paper.
It is a pleasure to thank Isabella Baccarelli, Fabio Pitari, and Caterina Caravita for their support with the CINECA environment. This work is supported by the Agenzia Spaziale Italiana (ASI) under - Euclid-FASE D  Attivita' scientifica per la missione - Accordo attuativo ASI-INAF n. 2018-23-HH.0, 
by the National Recovery and Resilience Plan (NRRP), Mission 4,
Component 2, Investment 1.1, Call for tender No. 1409 published on
14.9.2022 by the Italian Ministry of University and Research (MUR),
funded by the European Union – NextGenerationEU– Project Title
"Space-based cosmology with Euclid: the role of High-Performance
Computing" – CUP J53D23019100001 - Grant Assignment Decree No. 962
adopted on 30/06/2023 by the Italian Ministry of Ministry of
University and Research (MUR);
by the Italian Research Center on High-Performance Computing Big Data and Quantum Computing (ICSC), a project funded by the European Union - NextGenerationEU - and National Recovery and Resilience Plan (NRRP) - Mission 4 Component 2, by the INFN INDARK PD51 grant, and by the PRIN 2022 project EMC2 - Euclid Mission Cluster Cosmology: unlock the full cosmological utility of the Euclid photometric cluster catalog (code no. J53D23001620006). AF acknowledges support by the Excellence Cluster ORIGINS, which is funded by the Deutsche Forschungsgemeinschaft (DFG, German Research Foundation) under Germany’s Excellence Strategy - EXC-2094 - 390783311. AF acknowledges support from the Ludwig-Maximilians-Universit\"at in Munich.
The simulations were run partially on the Leonardo-Booster supercomputer as part of the Leonardo Early Access Program (LEAP) and under the ISCRA initiative. We acknowledge the CINECA award for the availability of high-performance computing resources and support. We acknowledge the use of the HOTCAT computing infrastructure of the Astronomical Observatory of Trieste -- National Institute for Astrophysics (INAF, Italy) \citep[see][]{2020ASPC..527..303B,2020ASPC..527..307T}.
\end{acknowledgements}

\bibliography{mybib}

\begin{thebibliography}{54}
\expandafter\ifx\csname natexlab\endcsname\relax\def\natexlab#1{#1}\fi

\bibitem[{Allen {et~al.}(2011)Allen, Evrard, \& Mantz}]{Allen:2011zs}
Allen, S.~W., Evrard, A.~E., \& Mantz, A.~B. 2011, ARA\&A, 49, 409

\bibitem[{Benson {et~al.}(2014)Benson, Ade, Ahmed, {et~al.}}]{SPT-3G:2014dbx}
Benson, B.~A., Ade, P. A.~R., Ahmed, Z., {et~al.} 2014, Proc. SPIE Int. Soc.
  Opt. Eng., 9153, 91531P

\bibitem[{Berner {et~al.}(2022)Berner, Refregier, Sgier, Kacprzak, Tortorelli,
  \& Monaco}]{Berner:2021hmp}
Berner, P., Refregier, A., Sgier, R., {et~al.} 2022, JCAP, 11, 002

\bibitem[{{Bertocco} {et~al.}(2020){Bertocco}, {Goz}, {Tornatore}, {Ragagnin},
  {Maggio}, {Gasparo}, {Vuerli}, {Taffoni}, \&
  {Molinaro}}]{2020ASPC..527..303B}
{Bertocco}, S., {Goz}, D., {Tornatore}, L., {et~al.} 2020, in Astronomical
  Society of the Pacific Conference Series, Vol. 527, 303

\bibitem[{Bocquet {et~al.}(2019)Bocquet, Dietrich, Schrabback,
  {et~al.}}]{SPT:2018njh}
Bocquet, S., Dietrich, J.~P., Schrabback, T., {et~al.} 2019, ApJ, 878, 55

\bibitem[{{Borgani} {et~al.}(2001){Borgani}, {Rosati}, {Tozzi}, {Stanford},
  {Eisenhardt}, {Lidman}, {Holden}, {Della Ceca}, {Norman}, \&
  {Squires}}]{Borgani:2001}
{Borgani}, S., {Rosati}, P., {Tozzi}, P., {et~al.} 2001, \apj, 561, 13

\bibitem[{Castro {et~al.}(2020)Castro, Borgani, Dolag, Marra, Quartin, Saro, \&
  Sefusatti}]{Castro:2020yes}
Castro, T., Borgani, S., Dolag, K., {et~al.} 2020, MNRAS, 500, 2316

\bibitem[{Castro \& Fumagalli(2024)}]{Castro_CCToolkit_A_Python_2024}
Castro, T. \& Fumagalli, A. 2024, {CCToolkit: A Python Package for Cluster
  Cosmology Calculations}

\bibitem[{Chuang {et~al.}(2015)Chuang, Zhao, Prada, {et~al.}}]{Chuang:2014toa}
Chuang, C.-H., Zhao, C., Prada, F., {et~al.} 2015, MNRAS, 452, 686

\bibitem[{Cohn(2006)}]{Cohn:2005ex}
Cohn, J.~D. 2006, New Astron., 11, 226

\bibitem[{Costanzi {et~al.}(2021)Costanzi, Saro, Bocquet,
  {et~al.}}]{DES:2020cbm}
Costanzi, M., Saro, A., Bocquet, S., {et~al.} 2021, PRD, 103, 043522

\bibitem[{Davis {et~al.}(1985)Davis, Efstathiou, Frenk, \&
  White}]{Davis:1985rj}
Davis, M., Efstathiou, G., Frenk, C.~S., \& White, S. D.~M. 1985, ApJ, 292, 371

\bibitem[{{DESI Collaboration: {Aghamousa}} {et~al.}(2016){DESI Collaboration:
  {Aghamousa}}, Aguilar, Ahlen, {et~al.}}]{DESI:2016fyo}
{DESI Collaboration: {Aghamousa}}, A., Aguilar, J., Ahlen, S., {et~al.} 2016
  [\eprint[arXiv]{1611.00036}]

\bibitem[{Dolag {et~al.}(2009)Dolag, Borgani, Murante, \&
  Springel}]{Dolag:2008ar}
Dolag, K., Borgani, S., Murante, G., \& Springel, V. 2009, MNRAS, 399, 497

\bibitem[{{Euclid Collaboration:}~Castro {et~al.}(2023){Euclid
  Collaboration:}~Castro, Fumagalli, Angulo, {et~al.}}]{Euclid:2022dbc}
{Euclid Collaboration:}~Castro, T., Fumagalli, A., Angulo, R.~E., {et~al.}
  2023, A\&A, 671, A100

\bibitem[{{Euclid Collaboration:}~Castro {et~al.}(2024){Euclid
  Collaboration:}~Castro, Fumagalli, Angulo, {et~al.}}]{Euclid:2024wog}
{Euclid Collaboration:}~Castro, T., Fumagalli, A., Angulo, R.~E., {et~al.}
  2024, A\&A, 691, A62

\bibitem[{{Euclid Collaboration:}~Fumagalli {et~al.}(2024){Euclid
  Collaboration:}~Fumagalli, Saro, Borgani, {et~al.}}]{Euclid:2022txd}
{Euclid Collaboration:}~Fumagalli, A., Saro, A., Borgani, S., {et~al.} 2024,
  A\&A, 683, A253

\bibitem[{{Euclid Collaboration:}~Mellier {et~al.}(2024){Euclid
  Collaboration:}~Mellier, Abdurro'uf, Acevedo~Barroso,
  {et~al.}}]{euclidoverview}
{Euclid Collaboration:}~Mellier, Y., Abdurro'uf, Acevedo~Barroso, J.~A.,
  {et~al.} 2024, A\&A, in press (Euclid SI), arXiv:2405.13491

\bibitem[{Fumagalli {et~al.}(2022)Fumagalli, Biagetti, Saro, Sefusatti, Slosar,
  Monaco, \& Veropalumbo}]{Fumagalli:2022plg}
Fumagalli, A., Biagetti, M., Saro, A., {et~al.} 2022, JCAP, 12, 022

\bibitem[{Fumagalli {et~al.}(2024)Fumagalli, Costanzi, Saro, Castro, \&
  Borgani}]{Fumagalli:2023yym}
Fumagalli, A., Costanzi, M., Saro, A., Castro, T., \& Borgani, S. 2024, A\&A,
  682, A148

\bibitem[{Fumagalli {et~al.}(2021)Fumagalli, Saro, Borgani,
  {et~al.}}]{Euclid:2021api}
Fumagalli, A., Saro, A., Borgani, S., {et~al.} 2021, A\&A, 652, A21

\bibitem[{Hitz {et~al.}(2024)Hitz, Berner, Crichton, Hennig, \&
  Refregier}]{Hitz:2024cwl}
Hitz, P., Berner, P., Crichton, D., Hennig, J., \& Refregier, A. 2024
  [\eprint[arXiv]{2410.01694}]

\bibitem[{Hu \& Kravtsov(2003)}]{Hu:2002we}
Hu, W. \& Kravtsov, A.~V. 2003, ApJ, 584, 702

\bibitem[{Kravtsov \& Borgani(2012)}]{Kravtsov:2012zs}
Kravtsov, A. \& Borgani, S. 2012, Ann. Rev. A\&A, 50, 353

\bibitem[{Landy \& Szalay(1993)}]{Landy:1993yu}
Landy, S.~D. \& Szalay, A.~S. 1993, \apj, 412, 64

\bibitem[{Laureijs {et~al.}(2011)Laureijs, Amiaux, Arduini,
  {et~al.}}]{EUCLID:2011zbd}
Laureijs, R., Amiaux, J., Arduini, S., {et~al.} 2011, arXiv:1110.3193

\bibitem[{{LSST Science Collaborations:}~{Abell} {et~al.}(2009){LSST Science
  Collaborations:}~{Abell}, Allison, Anderson, {et~al.}}]{LSSTScience:2009jmu}
{LSST Science Collaborations:}~{Abell}, P.~A., Allison, J., Anderson, S.~F.,
  {et~al.} 2009 [\eprint[arXiv]{0912.0201}]

\bibitem[{Maartens {et~al.}(2015)Maartens, Abdalla, Jarvis, \&
  Santos}]{Maartens:2015mra}
Maartens, R., Abdalla, F.~B., Jarvis, M., \& Santos, M.~G. 2015, PoS, AASKA14,
  016

\bibitem[{Majumdar \& Mohr(2004)}]{Majumdar:2003mw}
Majumdar, S. \& Mohr, J.~J. 2004, ApJ, 613, 41

\bibitem[{Mana {et~al.}(2013)Mana, Giannantonio, Weller, Hoyle, Huetsi, \&
  Sartoris}]{Mana:2013qba}
Mana, A., Giannantonio, T., Weller, J., {et~al.} 2013, MNRAS, 434, 684

\bibitem[{Marulli {et~al.}(2016)Marulli, Veropalumbo, \&
  Moresco}]{Marulli:2015jil}
Marulli, F., Veropalumbo, A., \& Moresco, M. 2016, Astron. Comput., 14, 35

\bibitem[{Marulli {et~al.}(2018)}]{Marulli:2018owk}
Marulli, F. {et~al.} 2018, A\&A, 620, A1

\bibitem[{Meiksin \& White(1999)}]{Meiksin:1998mu}
Meiksin, A. \& White, M.~J. 1999, MNRAS, 308, 1179

\bibitem[{Michaux {et~al.}(2021)Michaux, Hahn, Rampf, \&
  Angulo}]{Michaux:2020yis}
Michaux, M., Hahn, O., Rampf, C., \& Angulo, R.~E. 2021, MNRAS, 500, 663

\bibitem[{Monaco(2016)}]{Monaco:2016pys}
Monaco, P. 2016, Galaxies, 4, 53

\bibitem[{Monaco {et~al.}(2013)Monaco, Sefusatti, Borgani, Crocce, Fosalba,
  Sheth, \& Theuns}]{Monaco:2013qta}
Monaco, P., Sefusatti, E., Borgani, S., {et~al.} 2013, MNRAS, 433, 2389

\bibitem[{Monaco {et~al.}(2002)Monaco, Theuns, \& Taffoni}]{Monaco:2001jg}
Monaco, P., Theuns, T., \& Taffoni, G. 2002, MNRAS, 331, 587

\bibitem[{Munari {et~al.}(2017)Munari, Monaco, Sefusatti, Castorina, Mohammad,
  Anselmi, \& Borgani}]{Munari:2016aut}
Munari, E., Monaco, P., Sefusatti, E., {et~al.} 2017, MNRAS, 465, 4658

\bibitem[{{Planck Collaboration XX:}~Ade {et~al.}(2014){Planck Collaboration
  XX:}~Ade, Aghanim, Armitage-Caplan, {et~al.}}]{Planck:2013lkt}
{Planck Collaboration XX:}~Ade, P. A.~R., Aghanim, N., Armitage-Caplan, C.,
  {et~al.} 2014, A\&A, 571, A20

\bibitem[{{Planck Collaboration XXIV:}~Ade {et~al.}(2016){Planck Collaboration
  XXIV:}~Ade, Aghanim, Arnaud, {et~al.}}]{Planck:2015lwi}
{Planck Collaboration XXIV:}~Ade, P. A.~R., Aghanim, N., Arnaud, M., {et~al.}
  2016, A\&A, 594, A24

\bibitem[{Predehl {et~al.}(2021)}]{eROSITA:2020emt}
Predehl, P. {et~al.} 2021, A\&A, 647, A1

\bibitem[{Sartoris {et~al.}(2016)Sartoris, Biviano, Fedeli,
  {et~al.}}]{Sartoris:2015aga}
Sartoris, B., Biviano, A., Fedeli, C., {et~al.} 2016, \mnras, 459, 1764

\bibitem[{Schuecker {et~al.}(2003)Schuecker, Bohringer, Collins, \&
  Guzzo}]{Schuecker:2002ti}
Schuecker, P., Bohringer, H., Collins, C.~A., \& Guzzo, L. 2003, A\&A, 398, 867

\bibitem[{Scoccimarro {et~al.}(1999)Scoccimarro, Zaldarriaga, \&
  Hui}]{Scoccimarro:1999kp}
Scoccimarro, R., Zaldarriaga, M., \& Hui, L. 1999, \apj, 527, 1

\bibitem[{Sellentin \& Starck(2019)}]{Sellentin:2019ehy}
Sellentin, E. \& Starck, J.-L. 2019, JCAP, 08, 021

\bibitem[{Spergel {et~al.}(2015)Spergel, Gehrels, Baltay,
  {et~al.}}]{Spergel:2015sza}
Spergel, D., Gehrels, N., Baltay, C., {et~al.} 2015
  [\eprint[arXiv]{1503.03757}]

\bibitem[{Springel(2005)}]{Springel:2005mi}
Springel, V. 2005, MNRAS, 364, 1105

\bibitem[{Springel {et~al.}(2021)Springel, Pakmor, Zier, \&
  Reinecke}]{Springel:2020plp}
Springel, V., Pakmor, R., Zier, O., \& Reinecke, M. 2021, MNRAS, 506, 2871

\bibitem[{Springel {et~al.}(2001)Springel, White, Tormen, \&
  Kauffmann}]{Springel:2000qu}
Springel, V., White, S. D.~M., Tormen, G., \& Kauffmann, G. 2001, MNRAS, 328,
  726

\bibitem[{{Taffoni} {et~al.}(2020){Taffoni}, {Becciani}, {Garilli}, {Maggio},
  {Pasian}, {Umana}, {Smareglia}, \& {Vitello}}]{2020ASPC..527..307T}
{Taffoni}, G., {Becciani}, U., {Garilli}, B., {et~al.} 2020, in Astronomical
  Society of the Pacific Conference Series, Vol. 527, 307

\bibitem[{Takada \& Hu(2013)}]{Takada:2013wfa}
Takada, M. \& Hu, W. 2013, PRD, 87, 123504

\bibitem[{To {et~al.}(2021)}]{DES:2020uce}
To, C.-H. {et~al.} 2021, MNRAS, 502, 4093

\bibitem[{Vikhlinin {et~al.}(2009)Vikhlinin, Kravtsov, Burenin,
  {et~al.}}]{Vikhlinin:2008ym}
Vikhlinin, A., Kravtsov, A.~V., Burenin, R.~A., {et~al.} 2009, ApJ, 692, 1060

\bibitem[{Watson {et~al.}(2013)Watson, Iliev, D'Aloisio, Knebe, Shapiro, \&
  Yepes}]{Watson:2012mt}
Watson, W.~A., Iliev, I.~T., D'Aloisio, A., {et~al.} 2013, MNRAS, 433, 1230

\end{thebibliography}

\end{document}